\documentclass[conference]{IEEEtran}
\IEEEoverridecommandlockouts
\usepackage{cite}
\usepackage{amsmath,amssymb,amsfonts}
\usepackage{algorithm}
\usepackage{algorithmic}
\usepackage{graphicx}
\usepackage{textcomp}
\usepackage{xcolor}
\usepackage{caption}
\usepackage{subcaption}
\usepackage{comment}
\usepackage{multirow}
\usepackage{diagbox}
\usepackage{balance}

\usepackage{amsthm}

\usepackage{enumitem}

\usepackage{color, colortbl}
\definecolor{Gray}{gray}{0.9}
\definecolor{LightCyan}{rgb}{0.88,1,1}

\newcolumntype{a}{>{\columncolor{Gray}}c}
\newcolumntype{b}{>{\columncolor{LightCyan}}c}
\def\BibTeX{{\rm B\kern-.05em{\sc i\kern-.025em b}\kern-.08em
    T\kern-.1667em\lower.7ex\hbox{E}\kern-.125emX}}

\usepackage{pifont}% http://ctan.org/pkg/pifont

\begin{document}

\title{ROME: A Multi-Resource Job Scheduling Framework for Exascale HPC Systems}

\author{\IEEEauthorblockN{Yuping Fan}
\IEEEauthorblockA{\textit{Illinois Institute of Technology} \\
\textit{Chicago, IL}\\
yfan22@hawk.iit.edu}
}

\maketitle

\begin{abstract}
High-performance computing (HPC) is undergoing significant changes. Next generation HPC systems are equipped with diverse global and local resources, such as I/O burst buffer resources, memory resources (e.g., on-chip and off-chip RAM, external RAM/NVRA), network resources, and possibly other resources. Job schedulers play a crucial role in efficient use of resources. However, traditional job schedulers are single-objective and fail to efficient use of other resources. In this paper, we propose ROME, a novel multi-dimensional job scheduling framework to explore potential tradeoffs among multiple resources and provides balanced scheduling decision. Our design leverages genetic algorithm as the multi-dimensional optimization engine to generate fast scheduling decision and to support effective resource utilization.
\end{abstract}


\section{Introduction}
HPC systems and applications are undergoing dramatic changes. To ensure HPC systems meet diverse science and engineering application demands, increasingly diverse kinds of resources are added to the next generation HPC systems, such as I/O burst buffer resources \cite{LiuBB}, memory resources, and network resources \cite{Qiao1, Qiao2, Li1}. In order to absorb applications’ increasingly intensive and concurrent I/O requests, many next generation HPC systems are equipped with burst buffer served as the intermediate storage layer between compute nodes and relative slow parallel file system (PFS) \cite{LiuBB}. Trinity \cite{Trinity} at Los Alamos National Laboratory (LANL) \cite{LANL} and Cori \cite{Cori} at the National Energy Research Scientific Computing Center (NERSC) \cite{NERSC} introduce global burst buffer. Similarly, Theta \cite{Theta} at Argonne Leadership Computing Facility (ALCF) \cite{ALCF} and Summit \cite{Summit} at Oak Ridge Leadership Computing Facility (OLCF) \cite{OLCF} are equipped with local burst buffer (SSD-on-node). Shared memory resources are also incorporated in HPC systems to reduce the memory access latency of memory-hungry applications. Cooley \cite{Cooley} at ALCF augments its data analysis and visualization capability by installing shared memory.

Despite the rapid changes in HPC systems, the current job schedulers are not keeping up with these changes. The existing job schedulers are single-objective and fail to utilize diverse resources effectively in HPC systems. To address this problem, we propose ROME, a multi-dimensional job scheduling frame- work exploring the multi-objective formulation to solve the multi-dimensional job scheduling problem. ROME consists of a window-based mechanism for preserving job fairness, a multi-dimensional job scheduler for generating a set of solutions optimizing utilization of multiple resources via genetic algorithm, a decision maker for selecting single preferred solution from a set of optimal solutions. The extensive trace-based simulations \cite{CQSim, Fan6, CQSimGithub} on Mira logs \cite{Fan2} demonstrated that our framework is capable of improving resource utilization by up to 20\%, while reducing average job wait time.

\section{Methodology}\label{Methodology}

\begin{figure*}[htbp]
\centering
\centerline{\includegraphics[width=\linewidth]{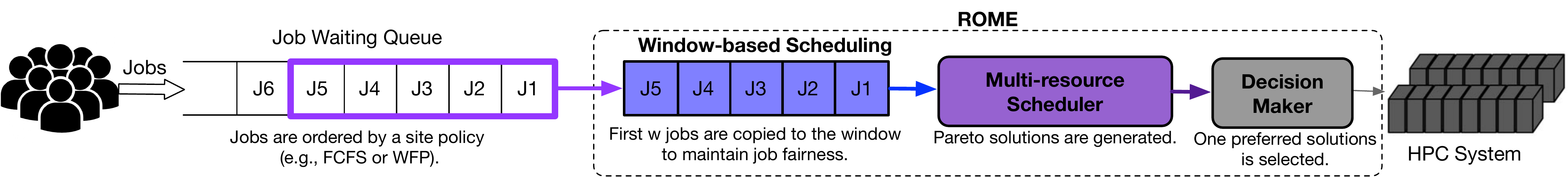}}
\caption{The overview of our multi-dimensional batch scheduling framework.}
\label{system_overview}
\end{figure*}

Figure \ref{system_overview} depicts a high-level overview of our multi-dimensional job scheduling framework. Our framework consists of three main components: a window-based mechanism, a multi-dimensional job scheduler, and a decision maker. Every job submission or job end event triggers a scheduling instance. At each scheduling instance, the original job scheduling policy in a system first orders jobs in the waiting queue and then we copy the first $w$ jobs to the window. Once jobs enter the window, the job scheduler first retrieves the resource utilization information, such as utilization of compute, SSD and burst buffer resources, and then provides optimal scheduling solutions using the genetic algorithm solver. One preferred optimal scheduling solution is selected as the final scheduling decision by the decision maker based on the preference information provided by the system administrator.

\subsection{Window-based Mechanism}\label{Window-based Mechanism}
Striking the balance between job fairness and scheduling performance is the one of the most challenging aspects of scheduling \cite{Fan3, Fan4}. To address the challenge, window-based mechanism is used to select first $w$ jobs to enter a window after sorting waiting jobs by the original scheduling policy, e.g., FCFS or WFP \cite{Fan1, Topper}. This mechanism only allows jobs in the window to be scheduled. Thereby, it preserves the job ordering to some extends.

\subsection{Multi-dimensional Job Scheduler}\label{Multi-dimensional Job Scheduler}
In order to illustrate the multi-dimensional capability of our job scheduler, we use compute nodes and burst buffer as example. Suppose the system has $N$ nodes and $R$ GB burst buffer. At a scheduling instance, there are $N$ used number of nodes and $R$ GB burst buffer in use. The objectives of the scheduler are maximizing compute node utilization and maximizing burst buffer utilization with the constraints that the requested and used resources do not exceed the total resources in the system. This problem can be formalized as follows:

\begin{align}
\text{maximize} & \sum \limits_{i=1}^w n_i x_i \\
\text{maximize}  & \sum \limits_{i=1}^w r_i x_i  \\
\text{subject to} &\notag \\
& \sum \limits_{i=1}^w n_i x_i \le N-N_{used}\\
& \sum \limits_{i=1}^w r_i x_i \le R-R_{used}\\
& x_i \in \{0, 1\}
\end{align}

The main challenge of solving the above problem is that it need to be solved by less than 30 seconds \cite{Fan5}, so that it can be used in practice. However, it is impossible to be solved by exact approaches, because $2^w$ solutions need to be examined. Therefore, we adopt a stochastic approach, genetic algorithm (GA) \cite{Mitchell}, to solve this problem. The ability of running in parallel is another attractive characteristic of GA, which makes it suitable to solve the multi-dimensional scheduling problem. GA is motivated by natural selection process. Weak species are extinct by nature selection, while strong species survive to future generations. It introduces changes to genes in each generation by two operations: crossover and mutation.

\subsection{Decision Maker}
The output of the multi-dimensional job scheduler is a set of non-dominated solutions and thereby the decision maker has to choose one solution from them. In general, among all the objectives in the multi-dimensional job scheduling problem, maximizing compute resource utilization is more important than other objectives. Hence, we first select the solution that maximizes compute resource utilization. Second, we make trade-off between this solution and other solutions. We will replace this solution by another solution if another solution can make great improvement in the utilization of other resources, while only causing a small deterioration in the utilization of compute resource. For example, we can consider those solutions that decrease the utilization of compute nodes by less than 10\% compared with the selected solution and increases the utilization of other resources more than 40\%. Then, we can choose one solution from those solutions with the maximum improvement on the utilization of other resources.

\begin{comment}
\section*{Acknowledgment}
This work is supported in part by US National Science Foundation grants CNS-1717763, CCF-1618776 and U.S. Department of Energy, Office of Science, under contract DE-AC02-06CH11357. 
\end{comment}

\balance
%\bibliographystyle{unsrt}
%\bibliographystyle{plainurl}
%\bibliographystyle{plainnat}
%\bibliography{ipdps1.bib}
\bibliographystyle{unsrt}
\bibliography{ipdps1.bib}

\end{document}